\documentclass[aps,prl,showpacs,floatfix]{revtex4}
\usepackage{graphicx}
\usepackage{times}
\usepackage{nicefrac}
\usepackage{amsmath}
\usepackage{amsfonts}
\usepackage{amssymb}
\usepackage{amsthm}
\usepackage{epsf}
\usepackage{bm}
\usepackage{bbm}

\usepackage{dcolumn}
\newcolumntype{.}{D{x}{}{-1}}

\newcommand{\be}{\begin{eqnarray}}
\newcommand{\ee}{\end{eqnarray}}
\newcommand{\la}{\langle}
\newcommand{\ra}{\rangle}

\newcommand{\veps}{\varepsilon}

\newcommand{\rmd}{{\rm d}}

\newcommand{\aZ}{\alpha Z}

\newcommand{\muN}{\mu_N}
\begin{document}

\title{Test of many-electron QED effects in the hyperfine splitting of
heavy high-$Z$ ions}
\author{A. V. Volotka,$^{1,2}$ D. A. Glazov,$^{2}$ O. V. Andreev,$^{2}$
V. M. Shabaev,$^{2}$ I. I. Tupitsyn,$^{2}$ and G. Plunien$^{1}$}

\affiliation{
$^1$ Institut f\"ur Theoretische Physik, Technische Universit\"at Dresden,
Mommsenstra{\ss}e 13, D-01062 Dresden, Germany \\
$^2$ Department of Physics, St. Petersburg State University,
Oulianovskaya 1, Petrodvorets, 198504 St. Petersburg, Russia \\
}

\begin{abstract}
A rigorous evaluation of the two-photon exchange corrections to the hyperfine
structure in lithiumlike heavy ions is presented. As a result, the theoretical
accuracy of the specific difference between the hyperfine splitting values of
H- and Li-like Bi ions is significantly improved. This opens a possibility for
the stringent test of the many-electron QED effects on a few percent level in
the strongest electromagnetic field presently available in experiments.
\end{abstract}


\maketitle
%
%
Accurate measurements of the ground-state hyperfine structure performed in H-like
$^{209}$Bi, $^{165}$Ho, $^{185}$Re, $^{187}$Re, $^{207}$Pb, $^{203}$Tl, and $^{205}$Tl
\cite{klaft:1994:2425,crespo:1996:826,crespo:1998:879,seelig:1998:4824,beiersdorfer:2001:032506}
were intended to probe QED in the strong electromagnetic field generated by a
heavy nucleus. However, accurate calculations revealed that the uncertainty of the
predicted hyperfine splittings, which mainly originates from the nuclear
magnetization distribution correction (the Bohr-Weisskopf effect), is comparable
in magnitude with the QED correction, see e.g. Refs.~\cite{persson:1996:1433}.
Accordingly, a direct identification of QED effects on the hyperfine splitting
in heavy H-like ions appeared to be unfeasible. It was shown instead, that this
uncertainty can be significantly reduced in a specific difference of the hyperfine
splitting values of H- and Li-like ions with the same nucleus \cite{shabaev:2001:3959}:
$\Delta^\prime E = \Delta E^{(2s)} - \xi \Delta E^{(1s)}$,
where $\Delta E^{(1s)}$ and $\Delta E^{(2s)}$ are the hyperfine splittings of H- and
Li-like ions, respectively, and the parameter $\xi$ is chosen to cancel the
Bohr-Weisskopf correction. The parameter $\xi$ can be calculated to a rather
high accuracy independently of the employed nuclear magnetization distribution model.
Thereby, the stringent tests of QED in strong fields can be achieved by studying
the specific difference of the hyperfine splitting values in H- and Li-like ions.

Till recently there existed only an indirect measurement of the $2s$ hyperfine splitting
in lithiumlike Bi ion with an accuracy of about 3\% \cite{beiersdorfer:1998:3022}.
Direct measurements with high-precision laser spectroscopy are feasible at the
current experimental storage ring (ESR) and future HITRAP facilities in GSI
\cite{noertershaeuser:2011:131}. Just recently, after
13 years of various attempts, the hyperfine splitting of the ground state Li-like
Bi ion has been directly observed in GSI \cite{noertershaeuser:private}.

Achievement of the required theoretical accuracy for the specific hyperfine splitting
difference for H- and Li-like heavy ions demands the rigorous evaluation of various
QED and interelectronic-interaction effects. Since the influence of one-electron QED
corrections is considerably reduced in the specific difference, the total value of
$\Delta^\prime E$ is essentially determined by the screened radiative and
interelectronic-interaction corrections. Recently, the screened self-energy and a
dominant part of the screened vacuum-polarization contributions have been calculated
rigorously within a systematic QED approach \cite{volotka:2009:033005,glazov:2010:062112}.
This calculation represents an essential advance beyond the local screening potential
approximation employed in the previous works
\cite{sapirstein:2001:032506,glazov:2006:330,volotka:2008:062507}.
As concerns the interelectronic-interaction contribution, up to now it was evaluated
rigorously only up to the first order in $1/Z$, see, e.g., Ref.~\cite{shabaev:1998:149}.
The contributions of second- and higher-order in $1/Z$ were calculated within the Breit
approximation employing many-body perturbation theory and configuration-interaction methods
\cite{boucard:2000:59,zherebtsov:2000:701,sapirstein:2001:032506,volotka:2008:062507},
however, for high-$Z$ ions such calculations can provide only an approximate estimation
of these corrections. In the present Letter we report on the complete evaluation of the
second-order interelectronic-interaction corrections within a rigorous QED approach.
As the most interesting application of these results we present improved theoretical
predictions for the specific difference between the ground-state hyperfine splitting
values of H- and Li-like Bi ions.
%

\begin{figure}
\includegraphics[width=0.5\textwidth]{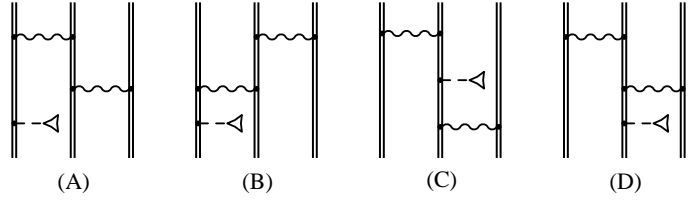}
\caption {Feynman diagrams representing the three-electron part of the
two-photon-exchange corrections to the hyperfine splitting. The wavy line
indicates the photon propagator and the double line indicates the electron
propagators in the Coulomb field. The dashed line terminated with the triangle
denotes the hyperfine interaction.}
\label{fig-3el}
\end{figure}

\begin{figure}
\includegraphics[width=0.5\textwidth]{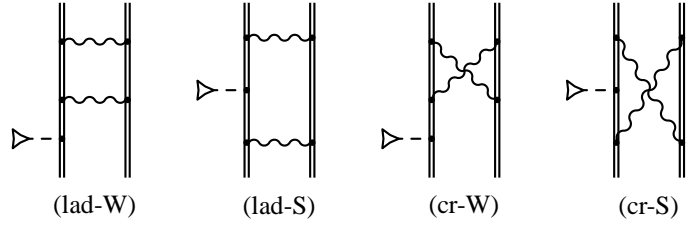}
\caption {Feynman diagrams representing the two-electron part of the
two-photon-exchange corrections to the hyperfine splitting. Notations are
the same as in Fig.~\ref{fig-3el}}
\label{fig-2el}
\end{figure}

The second-order interelectronic-interaction corrections in the presence of
an external potential correspond to the third-order perturbation theory terms.
Nowadays, several approaches are used for derivation of the formal expressions
for perturbation serial terms from the first principles of QED:
the two-time Green-function method \cite{shabaev:2002:119},
the covariant-evolution-operator method \cite{lindgren:2004:161},
and the line profile approach \cite{andreev:2008:135}.
Here, we employ the two-time Green-function method. To simplify the derivation
we specify the formalism regarding the closed shell electrons as belonging to
a redefined vacuum. In this way we have to consider all two-loop diagrams for the
valence electron in the presence of magnetic perturbation, i.e. 30 nonequivalent
diagrams. These diagrams encompass the second-order interelectronic-interaction
corrections, the one-electron two-loop, and the screened one-loop radiative
corrections. In our recent works \cite{volotka:2009:033005,glazov:2010:062112}
we have evaluated the screened radiative corrections. The generic types of the
second-order interelectronic-interaction diagrams, where we now focus on, are
depicted in Figs.~\ref{fig-3el},~\ref{fig-2el}.
The interelectronic-interaction corrections to the hyperfine splitting were usually
denoted by the terms $B(\aZ)/Z$ and $C(Z,\aZ)/Z^2$, see e.g.
Ref.~\cite{volotka:2008:062507}. The term $B(\aZ)/Z$ determines the
interelectronic-interaction correction of the first order in $1/Z$,
while the interelectronic-interaction corrections of second- and higher-orders
are denoted by the term $C(Z,\aZ)/Z^2$. Here, we isolate the terms of the second-
and higher-orders: $C(Z,\aZ)/Z^2 = C(\aZ)/Z^2 + D(Z,\aZ)/Z^3$, where $C(\aZ)/Z^2$
corresponds to the interelectronic-interaction corrections of the second-order
in $1/Z$ only (Figs.~\ref{fig-3el},~\ref{fig-2el}), and the term $D(Z,\aZ)/Z^3$
represents the third- and higher-order corrections in $1/Z$.
As in Ref.~\cite{shabaev:2002:119}, we call the parts of diagrams {\em reducible}
when an intermediate-state energy coincides with the reference-state energy and
{\em irreducible} otherwise. The irreducible parts of the three-electron diagrams
depicted in Fig.~\ref{fig-3el} yield the following contributions
\be
\label{Atype}
C^{\rm 3el, A} &=& 2 Z^2 G_a \sum_{b_1,b_2} \sum_{P,Q} (-1)^{P+Q} {\sum_n}'
 \Biggl\{ \frac{\la Pa Pb_2 | I(\Delta_{Paa}) | \xi_a n \ra
                \la n b_1 | I(\Delta_{b_1Qb_1}) | Qb_2 Qb_1 \ra}
               {\veps_{b_2}-\veps_n}
 \nonumber\\
       &+&\frac{\la Pb_2 Pa | I(\Delta_{Pb_2b_2}) | \xi_{b_2} n \ra
                \la n b_1 | I(\Delta_{b_1Qb_1}) | Qa Qb_1 \ra}
               {\veps_a-\veps_n}
        + \frac{\la Pb_2 Pb_1 | I(\Delta_{Pb_2b_2}) | \xi_{b_2} n \ra
                \la n a | I(\Delta_{aQa}) | Qb_1 Qa \ra}
               {\veps_{b_1}-\veps_n}
 \nonumber\\
       &+&\frac{\la Pa Pb_1 | I(\Delta_{Pab_2}) | \xi_{b_2} n \ra
                \la n b_2 | I(\Delta_{b_2Qb_1}) | Qa Qb_1 \ra}
               {\veps_a+\veps_{b_1}-\veps_{b_2}-\veps_n}
        + \frac12
          \frac{\la Pb_2 Pb_1 | I(\Delta_{Pb_2a}) | \xi_a n \ra
                \la n a | I(\Delta_{aQb_1}) | Qb_2 Qb_1 \ra}
               {\veps_{b_1}+\veps_{b_2}-\veps_a-\veps_n}
 \Biggr\}\,,
\ee
\be
\label{Btype}
C^{\rm 3el, B} &=& 2 Z^2 G_a \sum_{b_1,b_2} \sum_{P,Q} (-1)^{P+Q} {\sum_n}'
 \Biggl\{ \frac{\la \xi_{Pa} Pb_2 | I(\Delta_{Paa}) | a n \ra
                \la n b_1 | I(\Delta_{b_1Qb_1}) | Qb_2 Qb_1 \ra}
               {\veps_{b_2}-\veps_n}
 \nonumber\\
       &+&\frac{\la \xi_{Pb_2} Pa | I(\Delta_{Pb_2b_2}) | b_2 n \ra
                \la n b_1 | I(\Delta_{b_1Qb_1}) | Qa Qb_1 \ra}
               {\veps_a-\veps_n}
        + \frac{\la \xi_{Pb_2} Pb_1 | I(\Delta_{Pb_2b_2}) | b_2 n \ra
                \la n a | I(\Delta_{aQa}) | Qb_1 Qa \ra}
               {\veps_{b_1}-\veps_n}
 \nonumber\\
       &+&\frac{\la \xi_{Pa} Pb_1 | I(\Delta_{Pab_2}) | b_2 n \ra
                \la n b_2 | I(\Delta_{b_2Qb_1}) | Qa Qb_1 \ra}
               {\veps_a+\veps_{b_1}-\veps_{b_2}-\veps_n}
        + \frac12
          \frac{\la \xi_{Pb_2} Pb_1 | I(\Delta_{Pb_2a}) | a n \ra
                \la n a | I(\Delta_{aQb_1}) | Qb_2 Qb_1 \ra}
               {\veps_{b_1}+\veps_{b_2}-\veps_a-\veps_n}
 \Biggr\}\,,
\ee
\be
\label{Ctype}
C^{\rm 3el, C} &=& Z^2 G_a \sum_{b_1,b_2} \sum_{P,Q} (-1)^{P+Q} {\sum_{n_1,n_2}}'
 \Biggl\{ \frac{\la Pb_2 Pa | I(\Delta_{Pb_2b_2}) | b_2 n_1 \ra \la n_1 | T_0 | n_2 \ra
                \la n_2 b_1 | I(\Delta_{b_1Qb_1}) | Qa Qb_1 \ra}
               {(\veps_a-\veps_{n_1})(\veps_a-\veps_{n_2})}
 \nonumber\\
       &+&2
          \frac{\la Pb_2 Pb_1 | I(\Delta_{Pb_2b_2}) | b_2 n_1 \ra \la n_1 | T_0 | n_2 \ra
                \la n_2 a | I(\Delta_{aQa}) | Qb_1 Qa \ra}
               {(\veps_{b_1}-\veps_{n_1})(\veps_{b_1}-\veps_{n_2})}
 \nonumber\\
       &+&\frac{\la Pa Pb_1 | I(\Delta_{Pab_2}) | b_2 n_1 \ra \la n_1 | T_0 | n_2 \ra
                \la n_2 b_2 | I(\Delta_{b_2Qb_1}) | Qa Qb_1 \ra}
               {(\veps_a+\veps_{b_1}-\veps_{b_2}-\veps_{n_1})
                (\veps_a+\veps_{b_1}-\veps_{b_2}-\veps_{n_2})}
 \nonumber\\
       &+&\frac12
          \frac{\la Pb_2 Pb_1 | I(\Delta_{Pb_2a}) | a n_1 \ra \la n_1 | T_0 | n_2 \ra
                \la n_2 a | I(\Delta_{aQb_1}) | Qb_2 Qb_1 \ra}
               {(\veps_{b_1}+\veps_{b_2}-\veps_a-\veps_{n_1})
                (\veps_{b_1}+\veps_{b_2}-\veps_a-\veps_{n_2})}
 \Biggr\}\,,
\ee
\be
\label{Dtype}
C^{\rm 3el, D} &=& 2 Z^2 G_a \sum_{b_1,b_2} \sum_{P,Q} (-1)^{P+Q} {\sum_n}'
 \Biggl\{ \frac{\la Pa \xi_{Pb_2} | I(\Delta_{Paa}) | a n \ra
                \la n b_1 | I(\Delta_{b_1Qb_1}) | Qb_2 Qb_1 \ra}
               {\veps_{b_2}-\veps_n}
 \nonumber\\
       &+&\frac{\la Pb_2 \xi_{Pa} | I(\Delta_{Pb_2b_2}) | b_2 n \ra
                \la n b_1 | I(\Delta_{b_1Qb_1}) | Qa Qb_1 \ra}
               {\veps_a-\veps_n}
        + \frac{\la Pb_2 \xi_{Pb_1} | I(\Delta_{Pb_2b_2}) | b_2 n \ra
                \la n a | I(\Delta_{aQa}) | Qb_1 Qa \ra}
               {\veps_{b_1}-\veps_n}
 \nonumber\\
       &+&\frac{\la Pa \xi_{Pb_1} | I(\Delta_{Pab_2}) | b_2 n \ra
                \la n b_2 | I(\Delta_{b_2Qb_1}) | Qa Qb_1 \ra}
               {\veps_a+\veps_{b_1}-\veps_{b_2}-\veps_n}
        + \frac12
          \frac{\la Pb_2 \xi_{Pb_1} | I(\Delta_{Pb_2a}) | a n \ra
                \la n a | I(\Delta_{aQb_1}) | Qb_2 Qb_1 \ra}
               {\veps_{b_1}+\veps_{b_2}-\veps_a-\veps_n}
 \Biggr\}\,,
\ee
where the prime on the sums over intermediate states indicates that terms
with vanishing denominators should be omitted in the summation.
The irreducible parts of the two-electron diagrams depicted in
Fig.~\ref{fig-2el} yield
\be
\label{ladWtype}
C^{\rm 2el, lad-W} = Z^2 G_a \sum_b \sum_{P,Q} (-1)^{P+Q}
 \frac{i}{\pi} \int_{-\infty}^\infty \rmd\omega {\sum_{n_1,n_2}}'\,
 \frac{\la Pa Pb | I(\omega) | n_1 n_2 \ra
       \la n_1 n_2 | I(\omega+\Delta_{PaQa}) | \xi_{Qa} Qb \ra}
      {(\veps_{Pa}+\omega-u\veps_{n_1})(\veps_{Qb}-\omega-\Delta_{PaQa}-u\veps_{n_2})}\,,
\ee
\be
\label{crWtype}
C^{\rm 2el, cr-W} = Z^2 G_a \sum_b \sum_{P,Q} (-1)^{P+Q}
 \frac{i}{\pi} \int_{-\infty}^\infty \rmd\omega {\sum_{n_1,n_2}}'\,
 \frac{\la Pa n_2 | I(\omega) | n_1 Qb \ra
       \la \xi_{Pb} n_1 | I(\omega-\Delta_{PaQa}) | n_2 Qa \ra}
      {(\veps_{Pa}-\omega-u\veps_{n_1})(\veps_{Qb}-\omega-u\veps_{n_2})}\,,
\ee
\be
\label{ladStype}
C^{\rm 2el, lad-S} &=& Z^2 G_a \sum_b \sum_{P,Q} (-1)^{P+Q}
 \frac{i}{2\pi} \int_{-\infty}^\infty \rmd \omega
 \nonumber \\
 &\times&{\sum_{n_1,n_2,n_3}}'\,
 \frac{\la Pa Pb | I(\omega) | n_1 n_2 \ra \la n_2 | T_0 | n_3 \ra
       \la n_1 n_3 | I(\omega+\Delta_{PaQa}) | Qa Qb \ra}
      {(\veps_{Pa}+\omega-u\veps_{n_1})(\veps_{Qb}-\omega-\Delta_{PaQa}-u\veps_{n_2})
       (\veps_{Qb}-\omega-\Delta_{PaQa}-u\veps_{n_3})}\,,
\ee
\be
\label{crStype}
C^{\rm 2el, cr-S} = Z^2 G_a \sum_b \sum_{P,Q} (-1)^{P+Q}
 \frac{i}{2\pi} \int_{-\infty}^\infty \rmd\omega {\sum_{n_1,n_2,n_3}}'\,
 \frac{\la Pa n_2 | I(\omega) | n_1 Qb \ra \la n_3 | T_0 | n_2 \ra
       \la Pb n_1 | I(\omega-\Delta_{PaQa}) | n_3 Qa \ra}
      {(\veps_{Pa}-\omega-u\veps_{n_1})(\veps_{Qb}-\omega-u\veps_{n_2})
       (\veps_{Qb}-\omega-u\veps_{n_3})}\,,
\ee
where the prime on the sums indicates that in the summation we omit the
reducible and infrared-divergent terms, namely, those
with $\veps_{n_1}+\veps_{n_2}=\veps_a+\veps_b$
in the ladder-W diagrams,
with $\veps_{n_1}=\veps_{Pa},\,\veps_{n_2}=\veps_{Qb}$
in the direct parts of the cross-W diagrams,
with $\veps_{n_1}=\veps_{n_2}=\veps_a,\veps_b$
in the exchange parts of the cross-W diagrams,
with $\veps_{n_1}+\veps_{n_2}=\veps_a+\veps_b$ and
     $\veps_{n_1}+\veps_{n_3}=\veps_a+\veps_b$ and
     $\veps_{n_2}=\veps_{n_3}=\veps_{Qb}-\Delta_{PaQa}$
in the ladder-S diagrams,
with $\veps_{n_1}=\veps_{Pa},\,\veps_{n_2}=\veps_{Qb}$ and
     $\veps_{n_1}=\veps_{Pa},\,\veps_{n_3}=\veps_{Qb}$ and
     $\veps_{n_2}=\veps_{n_3}=\veps_{Qb}$
in the direct parts of the cross-S diagrams,
with $\veps_{n_1}=\veps_{n_2}=\veps_a,\veps_b$ and
     $\veps_{n_1}=\veps_{n_3}=\veps_a,\veps_b$ and
     $\veps_{n_2}=\veps_{n_3}=\veps_a,\veps_b$
in the exchange parts of the cross-S diagrams.
In Eqs.~(\ref{Atype})-(\ref{crStype}), $a$ and $b$ refer to the valence- and core-electron
states, respectively; the sum over $b$ runs over all closed-shell states, $P$ and $Q$ are
the permutation operators giving rise to the signs $(-1)^P$ and $(-1)^Q$ of the permutation,
respectively. $I(\omega)$ is the interelectronic-interaction operator \cite{shabaev:2002:119}, $u=1-i0$ preserves
the proper treatment of poles of the electron propagators. The energy difference
$\Delta_{n_1 n_2}$ is defined as $\Delta_{n_1 n_2} = \veps_{n_1} - \veps_{n_2}$.
$T_0$ is the electronic part of the hyperfine-interaction operator and $G_a$ is the
multiplicative factor depending on the quantum numbers of the valence electron (see for
details Ref.~\cite{volotka:2008:062507}). The modified wavefunction $|\xi\ra$ is defined
as follows
\be
\label{xi-oxi_a-1}
| \xi_a \ra = \sum_n^{\veps_n\neq\veps_a}
\frac{| n \ra \la n | T_0 | a \ra}{\veps_a-\veps_n}\,.
\ee
We refer to all contributions, where the energies of intermediate states and
the reference state coincide, the infrared-divergent contributions, and the
nondiagrammatic terms as the remaining reducible part.
Formal expressions for this part, which are rather bulky, will be published
elsewhere. Following the treatment of Ref.~\cite{shabaev:1994:4489} we introduce
a nonzero photon mass. In such a way we regularize the infrared divergences and
cancel them analytically.

Now let us discuss the numerical evaluation procedure. The infinite threefold
summations over the spectrum of the Dirac equation have been performed employing
the dual-kinetic-balance finite basis set method \cite{shabaev:2004:130405} with
basis functions constructed from B-splines \cite{sapirstein:1996:5213}. The Fermi
model for the nuclear charge density and the sphere model for the magnetic moment
distribution have been utilized. The summations over magnetic substates have been
performed analytically by means of standard formulas and also numerically as an
independent check. The most problematic part has consisted in evaluation of the
two-electron terms, which contain the integration over the energy of the virtual
photon $\omega$. In order to avoid strong oscillations arising for large real
values of $\omega$, we have performed a Wick rotation of the integration contour,
as it was done in the calculations of two-photon exchange corrections to the Lamb shift
Refs.~\cite{blundell:1993:2615,yerokhin:2000:4699,mohr:2000:052501},
and have employed the integration contours such as in Ref.~\cite{mohr:2000:052501}.
However, special care should be taken for the identification of the pole structure
of the integrands, because they are essentially more complicated than in the case
of the two-photon exchange corrections to the energy levels.

In what follows we present our result for the case of Li-like
Bi utilizing the following values for the nuclear properties:
$\la r^2 \ra^{1/2} = 5.5211$ fm \cite{angeli:2004:185},
$I^\pi=9/2-$, and $\mu=4.1106(2) \muN$ \cite{stone:2005:75}.
We have performed calculations in both Feynman and Coulomb gauges and the
corresponding individual contributions to the $C(\aZ)/Z^2$ are presented in
Table~\ref{tab:gauges}.
\begin{table}
\caption{Individual contributions to the two-photon exchange
correction $C(\aZ)/Z^2$ for the ground-state hyperfine
structure of the Li-like $^{209}$Bi$^{80+}$.}
\label{tab:gauges}
\tabcolsep3pt
\begin{tabular}{lrr}                      \hline
Contr.     &    Feynman  &    Coulomb  \\ \hline
3el, A     &    0.001685 &    0.002229 \\
3el, B     & $-$0.001942 & $-$0.002489 \\
3el, C     &    0.001154 &    0.001036 \\
3el, D     &    0.003781 &    0.003914 \\
2el, lad-W &    0.003401 &    0.003960 \\
2el, cr-W  &    0.000363 & $-$0.000019 \\
2el, lad-S &    0.001155 &    0.001226 \\
2el, cr-S  &    0.000207 & $-$0.000001 \\
reducible  & $-$0.009063 & $-$0.009116 \\ \hline
Total      &    0.000740 &    0.000740 \\ \hline
\end{tabular}
\end{table}
The total result is gauge independent on the level of the
numerical accuracy, that serves as an accurate check for both the derived
formulas and the numerical procedures. As an additional check, we
have reproduced the third-order many-body perturbation theory (MBPT)
result, considering the Breit approximation in the derived expressions.
Finally, we have found that the result of the rigorous QED evaluation of
the two-photon exchange correction $0.000740$ is in reasonable agreement
with the $1/Z^2$ term $0.00075$ extracted from the large-scale
configuration-interaction Dirac-Fock-Sturm calculation.

Now let us come to the consideration of the specific difference between
the ground state hyperfine splitting values for H-like and Li-like Bi. The
cancellation of the Bohr-Weisskopf effect appears with $\xi = 0.16886$
for the case of Bi.
In Table~\ref{tab:diff} we present the current status of individual
contributions to the specific difference. The rigorous evaluation of
the two-photon exchange corrections improves the accuracy of the
interelectronic-interaction term by an order of magnitude in comparison
with previous calculations \cite{volotka:2009:033005,glazov:2010:062112}.
The accuracy of the screened QED contribution has been also increased
due to recent rigorous evaluations of the Wichmann-Kroll parts of
the electric and magnetic loops \cite{andreev:2012:xxx}, which was
accounted for in Refs.~\cite{volotka:2009:033005,glazov:2010:062112}
within some approximations.
\begin{table}
\caption{Individual contributions to the specific
difference $\Delta^\prime E$ for $^{209}$Bi in meV.}
\label{tab:diff}
\begin{tabular}{lrrr}                                                                                   \hline
           &  $\Delta E^{(2s)}$\;\; &  $\xi \Delta E^{(1s)}$  & $\Delta^\prime E$\;\;\; \\ \hline
Dirac value                    &   844.829\phantom{(6)} &  876.638 & $-$31.809\phantom{(6)(3)} \\
Interelectronic interaction    &                                                               \\
$\sim 1/Z$                     & $-$29.995\phantom{(6)} &          & $-$29.995\phantom{(6)(3)} \\
$\sim 1/Z^2$                   &     0.258\phantom{(6)} &          &     0.258\phantom{(6)(3)} \\
$\sim 1/Z^3$ and higher orders &  $-$0.003(3)           &          &  $-$0.003(3)\phantom{(3)} \\
QED                            &  $-$5.052\phantom{(6)} & $-$5.088 &     0.036\phantom{(6)(3)} \\
Screened QED                   &     0.193(2)           &          &     0.193(2)\phantom{(3)} \\
Total                          &                        &          & $-$61.320(4)(5)    \\ \hline
\end{tabular}
\end{table}
Thus, the remaining theoretical uncertainty for the specific difference comes from the
uncalculated Wichmann-Kroll parts of the screened vacuum-polarization correction and
from the $1/Z^3$ and higher orders interelectronic-interaction term. The second
uncertainty in the total value of $\Delta^\prime E$ arises from the uncertainty of the
nuclear magnetic moment, the nuclear polarization corrections \cite{nefiodov:2003:35},
and other nuclear effects, which do not completely cancel in the specific difference.
It should be noted that the nuclear magnetic moment uncertainty can be larger due to
a chemical shift \cite{bastug:1996:281}. Employing the experimental value of the $1s$
hyperfine splitting $\Delta E^{(1s)}_{\rm exp} = 5.0840(8)$ eV \cite{klaft:1994:2425}
and the theoretical result for the specific difference, one can easily find the hyperfine
splitting for Li-like Bi $\Delta E^{(2s)} = 797.16(14)$ meV, where the accuracy is
fully determined by the uncertainty of the experimental value.
As one can see from Table~\ref{tab:diff}, the one-electron QED corrections are strongly
canceled in the specific difference and the dominant QED contributions comes from the
many-electron effects. Therefore, the theoretical accuracy achieved now for the specific
difference allows to test the many-electron QED effects on the few percent level,
provided the hyperfine splittings in H- and Li-like bismuth are measured with a
relative accuracy of about $10^{-6}$. When the QED corrections will be tested and
found to be valid, the comparison between the theoretical and experimental values will
enable the determination of the nuclear magnetic moments and their volume distribution.
%

In summary, we have rigorously calculated the two-photon exchange correction
to the hyperfine splitting in heavy Li-like ions. As a result we have significantly
increased the accuracy of the specific difference, thus providing the theoretical
prerequisite for a test of many-electron QED effects at strongest electromagnetic
fields. Further extensions of these calculations to the g factor of Li-like and
B-like heavy ions may also serve for an independent determination of the
fine structure constant from QED at strong fields \cite{shabaev:2006:253002}.
%

Valuable discussions with V. A. Yerokhin are gratefully acknowledged.
The work reported in this paper was supported
by the Deutsche Forschungsgemeinschaft (Grants No. VO 1707/1-1 and PL 254/7-1)
and GSI, by RFBR (Grant No. 10-02-00450), by the Russian Ministry of Education
and Science (Grant No. P1334), and by the grant of the President of the Russian
Federation (Grant No. MK-3215.2011.2). D.A.G. and O.V.A. acknowledge financial
support by the FAIR -- Russia Research Center, by the ``Dynasty'' foundation,
and by the G-RISC. Computing resources were provided by the Zentrum f\"ur
Informationsdienste und Hochleistungsrechnen (ZIH) at the TU Dresden.
%
%

%
\end{document}